\documentclass[aps,pra,reprint,superscriptaddress,floatfix]{revtex4-2}
\usepackage{amsmath,amssymb,bm,braket,graphicx}
\usepackage[colorlinks=true,linkcolor=blue,citecolor=blue,urlcolor=blue]{hyperref}

\begin{document}

\title{Finite-temperature mass gap and quench dynamics of mobile impurities in a Fermi gas}

\author{Baihua Gong}
\email{baihuagong@xjtu.edu.cn}
\affiliation{MOE Key Laboratory for Nonequilibrium Synthesis and Modulation of Condensed Matter, School of Physics, Xi'an Jiaotong University, Xi'an 710049,  China}

\date{\today}

\begin{abstract}
Recently, a mass-gap description of mobile impurities in a Fermi gas was introduced, which connects Anderson's orthogonality catastrophe for static impurities to the quasiparticle picture of Fermi polarons through a recoil-induced energy gap in the fermionic dispersion. That description, however, was restricted to zero temperature and did not address dynamics. Here we generalize the mass-gap model to finite temperature by combining the Lee--Low--Pines transformation with a self-consistent Hartree--Fock decoupling of the recoil-induced interaction, and we study the quench dynamics within this framework using the functional-determinant approach. At finite temperature the effective mass gap obeys the self-consistency equation $\Delta(T)=2U_F\tanh[\Delta(T)/4k_B T]$, with $U_F=k_F^2/2M$ and $M$ the impurity mass. This equation admits a nonzero solution below the characteristic temperature $T^*=U_F/(2k_B)$ and closes as $(T^*-T)^{1/2}$. We identify this closing as the mean-field signature of the thermal melting of the polaron and molecule quasiparticles. Computing the Ramsey response $S(t)$ after a sudden quench of the impurity--fermion interaction, we find that its long-time oscillations---quantum beats between the bound and in-gap states---disappear precisely above $T^*$. Our work ties the thermodynamic and dynamical fingerprints of polaron formation to a single temperature-dependent mean-field parameter.
\end{abstract}

\maketitle

\section{Introduction}
\label{sec:intro}

The problem of a mobile impurity immersed in a Fermi sea of host fermions is a paradigm of quantum many-body physics \cite{Massignan2014, Schmidt2018}. When the impurity is infinitely heavy and static, the ground state of the interacting system is orthogonal to the noninteracting Fermi sea in the thermodynamic limit---Anderson's orthogonality catastrophe (OC) \cite{Anderson1967,Nozieres1969}. For a mobile impurity of finite mass, by contrast, the impurity becomes dressed by the host fermions and forms a Fermi polaron, a quasiparticle characterized by a finite quasiparticle weight $Z$, an effective mass, and a finite lifetime \cite{Chevy2006,Combescot2007,Schirotzek2009}. Understanding how the quasiparticle picture of the polaron emerges from the OC as the impurity becomes mobile is a central challenge in this field.

A unified perspective on this question was recently provided by the mass-gap model \cite{Kain2017, Chen2025}. Applying the Lee--Low--Pines (LLP) transformation \cite{Lee1953} and reordering the recoil-induced interaction such that it vanishes on the Fermi sea, one obtains an effective mean-field Hamiltonian whose single-particle dispersion exhibits an energy gap $\Delta(M)=k_F^2/M$ at the Fermi momentum $k_F$. This recoil-induced gap acts as a regulator of the low-energy particle--hole excitations that underlie the OC, and it gives rise to an in-gap state that drives the polaron-to-molecule transition \cite{Punk2009,Schmidt2011}, as well as to a finite quasiparticle weight $Z$ \cite{Chen2025}. However, the mass-gap model has so far been confined to zero temperature and equilibrium, and thus cannot address the thermal melting of the polaron or its nonequilibrium dynamics. In this work we extend the model to both regimes.

To this end, we combine the Lee--Low--Pines transformation with a self-consistent Hartree--Fock decoupling of the recoil interaction, which yields the finite-temperature generalization of the operator reordering of Ref.~\cite{Chen2025}. We find that the resulting effective mass gap $\Delta(T)$ closes at a characteristic temperature $T^*$, signaling the thermal melting of the polaron and molecule quasiparticles, and that the Ramsey response $S(t)$, computed via the functional-determinant method \cite{Levitov1996,Klich2003}, loses its long-time oscillations precisely above $T^*$.

The remainder of this paper is organized as follows. In Sec.~\ref{sec:model} we introduce the model, derive the finite-temperature Hartree--Fock approximation, and analyze the resulting mass gap and characteristic temperature $T^*$. In Sec.~\ref{sec:quench} we present the quench protocol, the functional-determinant formalism, and the quench dynamics, including the infinite-mass and short-time benchmarks and the finite-temperature results. We conclude and discuss experimental implications in Secs.~\ref{sec:discussion} and \ref{sec:conclusion}.

\section{Model and finite-temperature mass-gap description}
\label{sec:model}

\subsection{Model}

We consider a single impurity of mass $M$ interacting with a noninteracting Fermi sea of $N$ fermions of mass $m$ through a contact interaction. The Hamiltonian reads
\begin{equation}
\hat H=\sum_{\bm k}\frac{k^2}{2M}\hat d^\dagger_{\bm k}\hat d_{\bm k}
+\sum_{\bm k}\frac{k^2}{2m}\hat c^\dagger_{\bm k}\hat c_{\bm k}
+\frac{g}{\mathcal V}\sum_{\bm k',\bm k,\bm q}\hat d^\dagger_{\bm k'+\bm q}\hat d_{\bm k'}\hat c^\dagger_{\bm k-\bm q}\hat c_{\bm k},
\label{eq:H}
\end{equation}
where $\hat d^{(\dagger)}_{\bm k}$ and $\hat c^{(\dagger)}_{\bm k}$ are the annihilation (creation) operators of the impurity and of the host fermions, respectively, and $\mathcal V$ is the  volume. The bare coupling $g$ is related to the physical scattering length $a$ through the standard regularization $1/g=m_r/(2\pi a)-m_r\Lambda/\pi^2$, with $m_r=mM/(m+M)$ the reduced mass and $\Lambda$ a momentum cutoff. We set $\hbar=1$ throughout and measure temperature in energy units ($k_B=1$ unless stated otherwise).

\subsection{Lee--Low--Pines transformation and self-consistent Hartree--Fock approximation}
\label{sec:HF}

We first consider the noninteracting limit of Eq.~\eqref{eq:H}, in which only the impurity and host kinetic energies are present. Applying the LLP transformation $\hat U=\exp\!\big(i\hat{\bm r}\cdot\sum_{\bm k}\bm k\,\hat c^\dagger_{\bm k}\hat c_{\bm k}\big)$ \cite{Lee1953} and working in the sector of zero total momentum ($P=0$), the impurity momentum operator is replaced by $\hat{\bm P}\to-\sum_{\bm k}\bm k\,\hat c^\dagger_{\bm k}\hat c_{\bm k}$, and the impurity kinetic energy becomes $\tfrac{1}{2M}\big(\sum_{\bm k}\bm k\,\hat c^\dagger_{\bm k}\hat c_{\bm k}\big)^2$. Normal-ordering with the fermionic anticommutation relations yields
\begin{equation}
\hat H_0
=\sum_{\bm k}\Bigl(\frac{k^2}{2m}+\frac{k^2}{2M}\Bigr)\hat c^\dagger_{\bm k}\hat c_{\bm k}
+\frac{1}{2M}\sum_{\bm k,\bm k'}(\bm k\cdot\bm k')\,\hat c^\dagger_{\bm k}\hat c^\dagger_{\bm k'}\hat c_{\bm k'}\hat c_{\bm k}.
\label{eq:H0_LLP}
\end{equation}
The first term is a renormalized free dispersion, while the second is the recoil-induced two-body interaction among the Fermi-sea particles. The latter was reordered in Refs.~\cite{Kain2017,Chen2025} at zero temperature such that it vanishes on the Fermi sea. We now generalize this construction to finite temperature by decoupling the recoil interaction in a self-consistent Hartree--Fock approximation.

In a normal (non-superfluid) state the only nonvanishing contraction is $\langle\hat c^\dagger_{\bm k}\hat c_{\bm k'}\rangle=n_{\bm k}\delta_{\bm k\bm k'}$. The direct (Hartree) terms vanish by the rotational symmetry of the Fermi sea, $\sum_{\bm k}\bm k\,n_{\bm k}=0$, while the exchange (Fock) terms contribute $-(1/M)\sum_{\bm k}k^2 n_{\bm k}\,\hat c^\dagger_{\bm k}\hat c_{\bm k}$. The Hartree--Fock Hamiltonian for the recoil part thus reads
\begin{equation}
\hat{\mathcal H}_0^{\rm HF}
=\sum_{\bm k}\Bigl[\frac{k^2}{2m}+\bigl(1-2n_{\bm k}\bigr)\frac{k^2}{2M}\Bigr]\hat c^\dagger_{\bm k}\hat c_{\bm k}
+\sum_{\bm k}\frac{k^2}{2M}\,n_{\bm k}^2,
\label{eq:H0_HF}
\end{equation}
where the last term is the double-counting correction ensuring $\langle\hat{\mathcal H}_0^{\rm HF}\rangle=\langle\hat H_0\rangle$.

At finite temperature the occupation numbers are the Fermi--Dirac distribution $n_{\bm k}=1/[e^{\beta(E_{\bm k}-\mu)}+1]$ with the single-particle dispersion
\begin{equation}
E_{\bm k}=\frac{k^2}{2m}+\bigl[1-2n_{\bm k}\bigr]\frac{k^2}{2M},
\label{eq:E_k}
\end{equation}
and the chemical potential $\mu$ fixed by particle-number conservation, $N=\sum_{\bm k}n_{\bm k}$. Equations \eqref{eq:E_k} together with the Fermi distribution and the number constraint form a closed set of self-consistency equations which are solved iteratively.

At zero temperature $n_{\bm k}=\theta(k_F-k)$, so that $1-2n_{\bm k}=-1$ for $k<k_F$ and $+1$ for $k>k_F$. Equation \eqref{eq:E_k} then reduces to
\begin{equation}
E_k=
\begin{cases}
\dfrac{k^2}{2m}-\dfrac{k^2}{2M}, & k<k_F,\\[6pt]
\dfrac{k^2}{2m}+\dfrac{k^2}{2M}, & k>k_F,
\end{cases}
\label{eq:mg_disp}
\end{equation}
which is precisely the mass-gap dispersion obtained by Kain and Ling \cite{Kain2017} and used in Ref.~\cite{Chen2025}. The self-consistent Hartree--Fock decoupling is thus the finite-temperature generalization of the operator-reordering procedure of Ref.~\cite{Chen2025}, and the two descriptions coincide at $T=0$.

The Hartree--Fock decoupling amounts to a Gaussian (quasifree) approximation of the interacting thermal state: the exact density matrix $e^{-\beta\hat H_0}/Z$ is replaced by the Gaussian state $e^{-\beta\hat{\mathcal H}_0^{\rm HF}}/Z$. In the quench setup considered below, the impurity--fermion interaction is switched off in the initial state ($g=0$ for $t<0$), so that the Gaussian approximation of the \emph{initial} density matrix involves only the recoil interaction, whereas the interaction $g$ is treated exactly as a sudden perturbation in the subsequent dynamics; we refer to this scheme as the \emph{recoil-only} Hartree--Fock approximation. It is distinct from the equilibrium Hartree--Fock treatment of Ref.~\cite{Kain2017}, in which $g$ enters the self-consistency loop and dresses the Hartree--Fock orbitals; the two approximations coincide only for the recoil-induced dispersion at $T=0$.

\subsection{Finite-temperature mass gap}
\label{sec:gap}

\begin{figure}[t]
    \centering
        \includegraphics[width=1\linewidth]{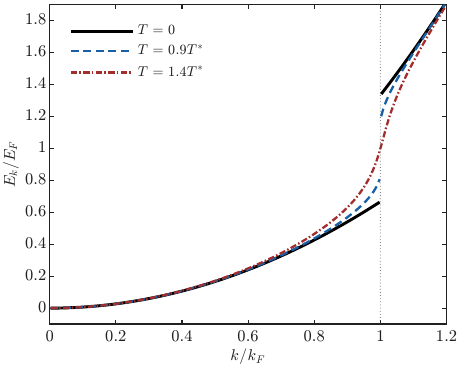}
        \includegraphics[width=1\linewidth]{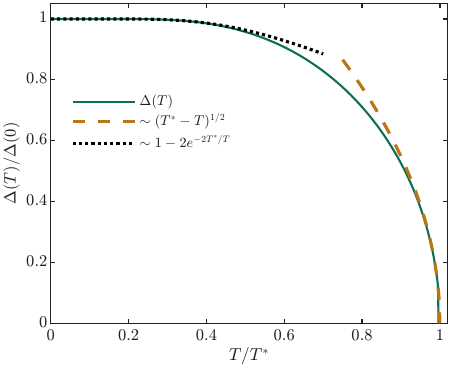}
\caption{ \label{fig:massgap}
(a) Mass-gap dispersion relation $E_k/E_F$ as a function of $k/k_F$ for $M/m=3$. At $T=0$ (solid) the dispersion exhibits a hard gap $\Delta(0)=k_F^2/M$ at the Fermi momentum; as the temperature increases toward $T^*$ (dashed) the gap shrinks, and for $T>T^*$ (dash-dotted) the dispersion is smooth and the gap has closed. (b) Effective mass gap $\Delta(T)/\Delta(0)$ as a function of $T/T^*$, obtained from Eq.~\eqref{eq:gap}. The gap equals $\Delta(0)=k_F^2/M$ at $T=0$ and closes at the characteristic temperature $T^*=U_F/2$;  the dotted line indicates the low-temperature behavior $ \Delta \sim 1 - 2 e^{-2 T^*/T}$ of Eq.~\eqref{eq:lowT}; the dashed line indicates the critical behavior $\Delta\sim(T^*-T)^{1/2}$ of Eq.~\eqref{eq:crit}.
    }
\end{figure}

We now analyze the self-consistency equations near the Fermi surface. Using the identity $1-2n_F(E)=\tanh[\beta(E-\mu)/2]$ the dispersion relation \eqref{eq:E_k} takes the form
\begin{equation}
E_{\bm k}=\varepsilon_k+U_k\tanh\!\Bigl(\frac{\beta(E_{\bm k}-\mu)}{2}\Bigr),
\label{eq:xi}
\end{equation}
with $\varepsilon_k=k^2/2m$ and $U_k=k^2/2M$.
At zero temperature the hyperbolic tangent reduces to a sign function, and Eq.~\eqref{eq:xi} admits the discontinuous solution $E_k=\varepsilon_k-U_k$ for $k<k_F$ and $E_k=\varepsilon_k+U_k$ for $k>k_F$, with a jump $\Delta(0)=2U_{k_F}=k_F^2/M$ at the Fermi momentum, in agreement with Eq.~\eqref{eq:mg_disp}. This gapped dispersion is illustrated in Fig.~\ref{fig:massgap}(a).

At any nonzero temperature the strict self-consistent solution of Eq.~\eqref{eq:xi} is smooth, since $n_{\bm k}$ is a smooth Fermi distribution, and the sharp jump is thermally smeared. To characterize the recoil-induced kink in the dispersion near the Fermi surface, we introduce an \emph{effective mass gap} $\Delta(T)$ within a step-dispersion (mean-field) ansatz: we assume that $E_{\bm k}$ jumps from $E_<$ to $E_>$ at $k_F$, with the chemical potential at the center of the jump, $\mu=(E_<+E_>)/2$, while the occupation retains its thermal Fermi--Dirac form $n_F(E_{\bm k}-\mu)$ evaluated on the jumped dispersion. Evaluating Eq.~\eqref{eq:xi} on the two sides of the jump, and using $1-2n_F(\mu\pm\Delta/2)=\pm\tanh(\beta\Delta/4)$ with $\Delta=E_>-E_<$, gives
$$
E_>=\varepsilon_{k_F}+U_{k_F}\tanh\!\Bigl(\frac{\beta\Delta}{4}\Bigr),
\qquad
E_<=\varepsilon_{k_F}-U_{k_F}\tanh\!\Bigl(\frac{\beta\Delta}{4}\Bigr).  
$$
Subtracting the two equations yields the closed self-consistency equation for the effective mass gap,
\begin{equation}
\Delta(T)=2U_F\tanh\!\Bigl(\frac{\Delta(T)}{4T}\Bigr).
\label{eq:gap}
\end{equation}
At zero temperature $\tanh(\infty)=1$, and we recover $\Delta(0)=2U_F=k_F^2/M$. Equation \eqref{eq:gap} is formally identical to the BCS gap equation; however, in contrast to BCS, the mass gap is a kinematic feature of the effective single-particle dispersion induced by the impurity recoil, rather than a symmetry-breaking order parameter. Accordingly, $\Delta(T)$ is a mean-field measure of the thermal smearing of the recoil-induced kink at the Fermi surface, and its closing at $T^*$ signals a smooth thermal crossover rather than a genuine phase transition.

Equation \eqref{eq:gap} always admits the trivial solution $\Delta=0$. A nonzero solution exists only for $T<T^*$, where the characteristic temperature $T^*$ is given by
\begin{equation}
  T^*=\frac{U_F}{2}=\frac{k_F^2}{4M}.
\label{eq:Tstar}
\end{equation}
For the representative mass ratio $M/m=3$ used in our numerical analysis, $T^*=E_F/6$ with $E_F=k_F^2/2m$. In the infinite-mass limit $M\to\infty$ the recoil energy $U_F$ vanishes, so that $\Delta(0)\to0$ and $T^*\to0$: the mass gap closes at all temperatures, and the static-impurity orthogonality catastrophe  is recovered in the zero-temperature limit. This is consistent with the OC being a genuine zero-temperature phenomenon that is not regularized in the static limit, where the recoil-induced mass gap is absent.

The gap equation \eqref{eq:gap} admits simple analytic expressions in two important limits.

\emph{Low-temperature limit ($T\ll T^*$).} Using $\tanh x\simeq1-2e^{-2x}$ for $x\gg1$, we obtain
\begin{equation}
\Delta(T)\simeq 2U_F\Bigl[1-2e^{-U_F/T}\Bigr]
=\Delta(0)\Bigl[1-2e^{- 2 T^* /T}+\cdots\Bigr].
\label{eq:lowT}
\end{equation}
The gap approaches its zero-temperature value exponentially, characteristic of a gapped spectrum with thermally activated excitations.

\emph{Critical region ($T\to T^{*-}$).} Writing $T=T^*(1-\delta)$ with $\delta\ll1$ and expanding $\tanh x\simeq x-x^3/3$ for $x\ll1$, Eq.~\eqref{eq:gap} gives
\begin{equation}
\Delta(T)\simeq 4\sqrt{3}\,T^*\,\sqrt{\,1-\frac{T}{T^*}\,},
\label{eq:crit}
\end{equation}
i.e., the effective gap closes continuously with the mean-field exponent $1/2$. The temperature dependence of $\Delta(T)$ obtained from Eq.~\eqref{eq:gap}, together with the two asymptotic limits \eqref{eq:lowT} and \eqref{eq:crit},  is shown in Fig.~\ref{fig:massgap}(b).

\section{Quench dynamics and Ramsey response}
\label{sec:quench}

\subsection{Quench protocol and functional determinant}
\label{sec:quench_protocol}

We now include the point-contact potential and consider a sudden quench in which the impurity--fermion interaction is switched on at $t=0$. For $t<0$ the system is in the thermal state of $\hat{\mathcal H}_0^{\rm HF}$ [Eq.~\eqref{eq:H0_HF}], i.e., with the interaction switched off, $g=0$. At $t=0$ the interaction is turned on, and for $t>0$ the system evolves under the quadratic Hamiltonian (up to the constant shift, which we drop)
\begin{equation}
\hat{\mathcal H}^{\rm HF}
=\sum_{\bm k}\Bigl[\frac{k^2}{2m}+\bigl(1-2n_{\bm k}\bigr)\frac{k^2}{2M}\Bigr]\hat c^\dagger_{\bm k}\hat c_{\bm k}
+\frac{g}{\mathcal V}\sum_{\bm k,\bm k'}\hat c^\dagger_{\bm k}\hat c_{\bm k'}.
\label{eq:H_HF}
\end{equation}
Throughout the quench the mean-field occupation $n_{\bm k}$ is kept frozen at its pre-quench self-consistent value, so that both the initial and final Hamiltonians are quadratic (single-particle) operators.

The central observable is the overlap (Ramsey response)
\begin{equation}
S(t)=\bigl\langle e^{i\hat{\mathcal H}_0^{\rm HF}t}e^{-i\hat{\mathcal H}^{\rm HF}t}\bigr\rangle,
\label{eq:S_def}
\end{equation}
where $\langle\cdots\rangle$ denotes the average in the initial thermal state. Because both Hamiltonians are quadratic, $S(t)$ reduces to a determinant over the single-particle orbitals \cite{Levitov1996,Klich2003},
\begin{equation}
S(t)=\det\!\bigl[ \mathbb I -\hat n+\hat n\,e^{i\hat h_0 t}e^{-i\hat h t}\bigr],
\label{eq:S_det}
\end{equation}
where $\hat h_0$ and $\hat h$ are the single-particle Hamiltonians corresponding to $\hat{\mathcal H}_0^{\rm HF}$ and $\hat{\mathcal H}^{\rm HF}$, respectively, and $\hat n$ is the occupation-number operator with eigenvalues $n_{\bm k}$. Equation \eqref{eq:S_det} is the standard Levitov--Klich formula and reduces to the form used in Ref.~\cite{Knap2012} in the infinite-mass limit.

The single-particle Hamiltonian underlying $\hat{\mathcal H}^{\rm HF}$ [Eq.~\eqref{eq:H_HF}] consists of the recoil-dressed continuum $E_{\bm k}$ together with a separable (rank-one) contact potential. The latter generates discrete states, whose energies $\omega$ are the solutions of the in-medium Lippmann--Schwinger condition
$$
\frac{1}{g}=\frac{1}{\mathcal V}\sum_{\bm k}\frac{1}{\omega-E_{\bm k}}.
$$
These solutions comprise a bound state below the bottom of the band and an in-gap state inside the recoil gap of Eq.~\eqref{eq:mg_disp}. Following Ref.~\cite{Chen2025}, the attractive polaron corresponds to occupying the bound state while leaving the in-gap state empty, whereas the molecule corresponds to occupying the in-gap state in addition to the bound state.

Numerically, exploiting rotational invariance, the rank-one (s-wave) structure of the contact potential reduces the Hamiltonian to a one-dimensional s-wave problem and the functional determinant in Eq.~\eqref{eq:S_det} to a scalar equation in the radial momentum, which greatly simplifies the evaluation. This equation is discretized on a uniform momentum grid, using the regularization of the contact interaction described in Sec.~\ref{sec:model}, and the results are checked for convergence with respect to the grid spacing and the momentum cutoff.

\subsection{Results}
\label{sec:results}

\begin{figure}[t]
\centering
\includegraphics[width=1\columnwidth]{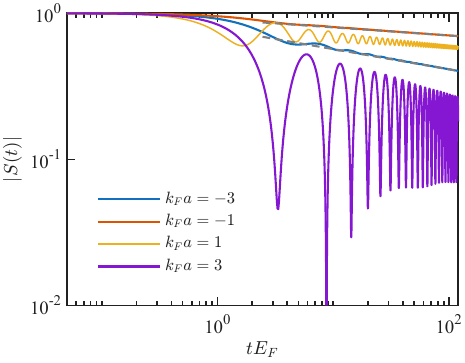}
\includegraphics[width=1\columnwidth]{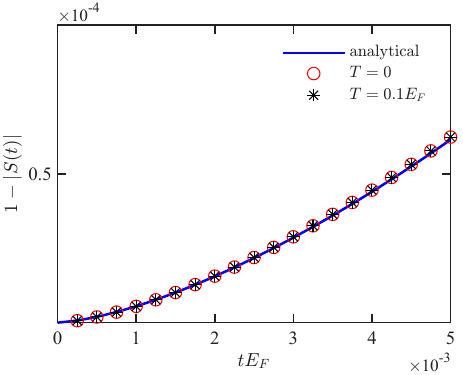}
\caption{\label{fig:validation}
Method validation. (a) Infinite-mass benchmark: Ramsey response $|S(t)|$ in the limit $M\to\infty$; the overlap decays as a power law $|S(t)|\sim t^{-(\delta_F/\pi)^2}$ and exhibits no finite quasiparticle weight. (b) Short-time behavior of $1-|S(t)|$ for $k_F a=3$ at $T=0$ and $T=0.1E_F$, compared with the universal analytic result $1-|S(t)| $ of Ref.~\cite{Parish2016} (solid line); the collapse of the numerical curves onto the analytic result at both temperatures validates the functional-determinant implementation and confirms the temperature independence of the short-time behavior.   }
\end{figure}

As a first check, in the infinite-mass limit $M\to\infty$ the recoil energy vanishes, the mass gap closes, and the dispersion reduces to the free one, so that Eq.~\eqref{eq:H_HF} becomes a free Fermi gas plus a static point-contact potential---the orthogonality-catastrophe setup of Ref.~\cite{Knap2012}. We have verified that our functional-determinant evaluation reproduces the results of Ref.~\cite{Knap2012} in this limit, including the power-law decay $|S(t)|\sim t^{-(\delta_F/\pi)^2}$ with $\delta_F$ the scattering phase shift at the Fermi momentum and the vanishing quasiparticle weight ($Z=0$) [Fig.~\ref{fig:validation}(a)]. As a second check, at short times $S(t)$ exhibits the universal, temperature-independent nonanalytic expansion
$$
S(t) \simeq 1 - \frac{8e^{-i\pi/4}(m/m_r)^{3/2}}{9\pi^{3/2}}\,(tE_F)^{3/2}.
$$
Since $|S(t)|$ remains close to unity at short times, we display $1-|S(t)|$ in Fig.~\ref{fig:validation}(b); our numerical results for $k_F a=3$ at $T=0$ and $T=0.1E_F$ both collapse onto the analytic result, confirming the temperature independence of the short-time behavior.

\begin{figure*}[t]
\centering
\includegraphics[width=0.32\textwidth]{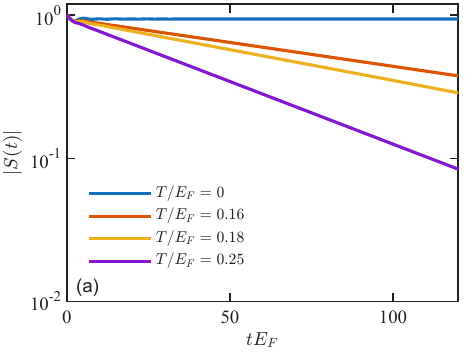}
\hfill
\includegraphics[width=0.32\textwidth]{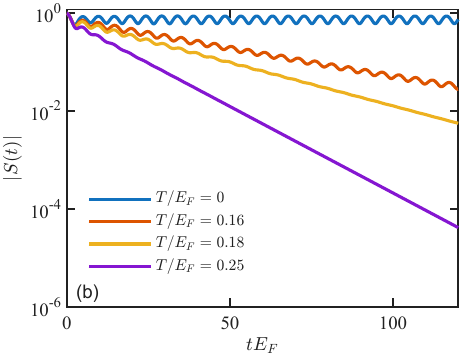}
 \hfill
\includegraphics[width=0.32\textwidth]{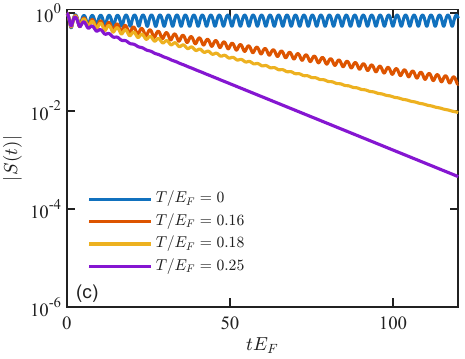}
\caption{\label{fig:quench}
Ramsey response $|S(t)|$ for finite mass ($M/m=3$) at the three representative interaction strengths (a)  $k_F a=-1$ (no bound state),(b) $k_F a=20$ (polaron) and (c) $k_F a=1$ (molecule). In each panel the four curves correspond to $T/E_F=0$, $0.16$, $0.18$, and $0.25$, straddling the characteristic temperature $T^*=E_F/6$. At $T=0$ the overlap saturates at long times to the quasiparticle weight $Z$, monotonically for $k_F a=-1$ and with oscillations for $k_F a=1$ and $k_F a=20$; at finite temperature $|S(t)|$ decays to zero. The long-time oscillations persist for $T<T^*$ ($T/E_F=0.16$) but disappear for $T>T^*$ ($T/E_F=0.18,0.25$), where $|S(t)|$ decays monotonically; the disappearance of the oscillations coincides with the closing of the mass gap at $T^*$.    }
\end{figure*}

We now present the overlap $S(t)$ obtained from Eq.~\eqref{eq:S_det} for the mass ratio $M/m=3$, for which $T^*=E_F/6$ [Eq.~\eqref{eq:Tstar}], and for three representative interaction strengths $k_F a=-1$, $1$, and $20$, corresponding respectively to the BCS side with no bound state (neither a polaron nor a molecule), the BEC side with a bound state and an occupied in-gap state (molecule), and the vicinity of unitarity with a bound state and an empty in-gap state (polaron) \cite{Chen2025}.

For finite mass at zero temperature, the mass gap $\Delta(0)=k_F^2/M$ regularizes the low-energy particle--hole excitations responsible for the orthogonality catastrophe, and the overlap saturates at long times to the quasiparticle weight
\begin{equation}
Z=\bigl|\braket{FS|\widetilde{FS}}\bigr|^2=\lim_{t\to\infty}|S(t)|,
\label{eq:Z}
\end{equation}
where $\ket{FS}$ and $\ket{\widetilde{FS}}$ denote the noninteracting Fermi sea and the ground state of $\hat{\mathcal H}^{\rm HF}$; this is to be contrasted with the infinite-mass limit, where $Z=0$. The discrete states imprint themselves on this saturation: at $T=0$ the approach to $Z$ is oscillatory when both a bound and an in-gap state are present, and monotonic when the bound state is absent, as shown by the $T=0$ curves in Fig.~\ref{fig:quench}.

At finite temperature the initial state is a thermal mixture, and $|S(t)|\to0$ at long times, in contrast to the nonzero zero-temperature quasiparticle weight $Z=\lim_{t\to\infty}|S(t)|$ of Eq.~\eqref{eq:Z}. The central result is the temperature dependence of the oscillations, shown in Fig.~\ref{fig:quench}: for $k_F a=1$ and $k_F a=20$ the long-time oscillations persist for $T<T^*$ ($T/E_F=0.16$) but disappear for $T>T^*$ ($T/E_F=0.18$ and $0.25$), where $|S(t)|$ decays monotonically, whereas for $k_F a=-1$ the decay is monotonic at all temperatures and shows no signature of $T^*$.

These oscillations are quantum beats between the two discrete states of $\hat{\mathcal H}^{\rm HF}$, namely the bound and in-gap states. At long times the overlap is dominated by the discrete contributions,
$$
S(t)\simeq Z_b\,e^{-i\omega_b t}+Z_{ig}\,e^{-i\omega_{ig}t}+S_{\rm cont}(t),
$$
where $\omega_b$ and $\omega_{ig}$ are the bound- and in-gap-state energies, $Z_b$ and $Z_{ig}$ their overlaps with the initial thermal state, and $S_{\rm cont}(t)$ the decaying continuum contribution. The modulus square thus oscillates at the difference of the two level energies,
$$
|S(t)|^2\simeq Z_b^2+Z_{ig}^2+2Z_b Z_{ig}\cos\!\bigl[(\omega_{ig}-\omega_b)t\bigr]+\cdots,
$$
i.e., at the quantum-beat frequency $\omega_{ig}-\omega_b$ rather than at either discrete energy alone. As the temperature increases, the effective mass gap $\Delta(T)$ closes according to Eq.~\eqref{eq:gap}, and the in-gap state is progressively absorbed into the thermally broadened continuum, so that its overlap $Z_{ig}\to0$; with only a single discrete level remaining, the beat---and hence the oscillation---disappears. The temperature $T^*$ at which this happens coincides with the closing of the mass gap, so that the thermodynamic melting of the quasiparticle and the disappearance of the dynamical oscillations are controlled by the same mean-field parameter $\Delta(T)$.

\section{Discussion}
\label{sec:discussion}

\emph{Nature of the transition.} The mass gap $\Delta(M)=k_F^2/M$ is a kinematic feature of the effective single-particle dispersion induced by the impurity recoil, not a symmetry-breaking order parameter. Accordingly, the closing of the gap at $T^*$ [Eq.~\eqref{eq:gap}] is the mean-field description of a smooth thermal crossover in which the polaron and molecule quasiparticles melt, rather than a genuine thermodynamic phase transition, in line with the smooth polaron-to-molecule crossover observed experimentally \cite{Ness2020}. Indeed, the strict self-consistent solution of Eq.~\eqref{eq:xi} is smooth at any $T>0$, and the effective gap $\Delta(T)$ emerges only within the step-dispersion (mean-field) ansatz of Sec.~\ref{sec:gap}. We therefore regard $T^*$ as a characteristic temperature that provides a transparent, analytically tractable estimate of the polaron melting scale, in qualitative agreement with the thermal broadening of the polaron spectral function found in diagrammatic approaches \cite{Hu2024a,Hu2024b}. The value of such a criterion is that it ties the melting to a simple, closed-form condition $\Delta(T^*)=0$ rather than to a detailed spectral calculation.

\emph{Validity of the approximation.} The functional-determinant result \eqref{eq:S_det} is exact for quadratic Hamiltonians, so the only approximation in our treatment is the Hartree--Fock (Gaussian) description of the initial thermal state together with its frozen mean field [Sec.~\ref{sec:HF}]. At zero temperature the residual recoil interaction neglected beyond Hartree--Fock vanishes on the Fermi sea \cite{Chen2025}, so the approximation is controlled at low temperature; the non-Gaussian correlations it generates are expected to grow with temperature and as the mass ratio $m/M\to1$, where quantum fluctuations of the recoil interaction are enhanced \cite{Chen2025}.

\emph{Experimental relevance.} The overlap $S(t)$ is directly measurable in Ramsey-interference experiments with mass-imbalanced cold-atom mixtures \cite{Knap2012,Cetina2016}. The predicted crossover from oscillatory to monotonic decay as a function of temperature provides a clear experimental fingerprint of the thermal melting of the polaron, and the characteristic temperature $T^*=k_F^2/(4Mk_B)$ sets the relevant scale. For typical mixtures (e.g., a $^{40}$K impurity immersed in a $^6$Li gas) this temperature is a sizable fraction of the Fermi temperature and is therefore within reach of current experiments.

\section{Conclusion}
\label{sec:conclusion}

We have extended the mass-gap description of mobile impurities in a Fermi gas to finite temperature and to nonequilibrium dynamics. At finite temperature the recoil-induced mass gap closes at a characteristic temperature $T^*$, signaling the thermal melting of the polaron and molecule quasiparticles. This melting has a direct dynamical counterpart: in the Ramsey response following a quench of the impurity--fermion interaction, the long-time oscillations, which are quantum beats between the discrete bound and in-gap states, disappear precisely above $T^*$. A single mean-field parameter $\Delta(T)$ thus governs both the equilibrium melting and the nonequilibrium dynamics, providing a compact, unified picture of how Fermi polarons lose their coherence with increasing temperature and connecting the mass-gap framework to the orthogonality-catastrophe physics of the static-impurity limit. Promising extensions include the residual recoil interaction beyond Hartree--Fock and the application of the mass-gap description to Bose gases \cite{Grusdt2025} and lower dimensions.

\section*{acknowledgments}
This work was supported by  the Special Funds  for Theoretical Physics  of the National Natural Science Foundation of China (Grant No. 11147117). We also thank the HPC platform of Xi'an Jiaotong University, where our numerical calculations were performed.

\section*{Data availability }
The data that support the findings of this article are not publicly available. The data are available from the authors upon reasonable request.

\end{document}